\documentclass{ecai}

\usepackage{amsmath}

\usepackage[belowskip=10pt]{caption}
\usepackage[belowskip=1pt]{subcaption}

\usepackage{soul}
\usepackage{url}
\usepackage{algorithm}
\usepackage{algorithmic}

\newtheorem{example}{Example}

\usepackage{multirow}%
\newcommand{\mpsHide}[1]{}
\usepackage[utf8]{inputenc}

\newcommand{\citepossessive}[1]{\citeauthor{#1}'s \cite{ #1}}

\usepackage{booktabs}
\usepackage{arydshln}
\makeatletter
\def\adl@drawiv#1#2#3{%
        \hskip.5\tabcolsep
        \xleaders#3{#2.5\@tempdimb #1{1}#2.5\@tempdimb}%
                #2\z@ plus1fil minus1fil\relax
        \hskip.5\tabcolsep}
\newcommand{\cdashlinelr}[1]{%
  \noalign{\vskip\aboverulesep
           \global\let\@dashdrawstore\adl@draw
           \global\let\adl@draw\adl@drawiv}
  \cdashline{#1}
  \noalign{\global\let\adl@draw\@dashdrawstore
           \vskip\belowrulesep}}
\makeatother

\usepackage{listings}
\newcommand{\framework}{\fsl{Exanna}\xspace}

\usepackage{xcolor}
\definecolor{revision}{rgb}{0.29, 0.0, 0.51}

\usepackage{siunitx}
\RequirePackage[np]{numprint}
\npdecimalsign{.}
\npthousandsep{,}

\RequirePackage{xspace}
\newcommand{\etal}{et al.\xspace}

\newcommand{\fsub}[1]{\textsubscript{#1}}

\newcommand{\fbf}{\textbf}

\newcommand{\fsl}{\textsl}

\DeclareMathAlphabet{\mathsl}{OT1}{ptm}{m}{sl}

\newcommand{\shyphen}{\text{--}} 

\usepackage{tikz}
\usepackage{pgfplots}
\pgfplotsset{compat=newest}
\usetikzlibrary{pgfplots.statistics}
\usepgfplotslibrary{groupplots}
\usepgfplotslibrary{statistics}

\usepgfplotslibrary{external}
\usepackage{pifont}
\newcommand{\cmark}{\ding{51}}%
\newcommand{\xmark}{\ding{55}}%
\newcommand{\InfectionRisk}{Risk}

\pgfplotsset{
    discard if not/.style 2 args={
        /pgfplots/boxplot/data filter/.code={
            \edef\tempa{\thisrow{#1}}
            \edef\tempb{#2}
            \ifx\tempa\tempb
            \else
                
            \fi
        }
    }
}

\begin{document}

\begin{frontmatter}

\paperid{2349}

\title{Value-Based Rationales Improve Social Experience: \\A Multiagent Simulation Study}

\author[A]{\fnms{Sz-Ting}~\snm{Tzeng}\orcid{0000-0001-9304-6566}}
\author[B]{\fnms{Nirav}~\snm{Ajmeri}\orcid{0000-0003-3627-097X}}
\author[C]{\fnms{Munindar P.}~\snm{Singh}\orcid{0000-0003-3599-3893}}

\address[A]{Ume{\aa} University, Ume{\aa}, Sweden}
\address[B]{University of Bristol, Bristol, UK}
\address[C]{North Carolina State University, Raleigh, NC, USA}

\begin{abstract}
We propose \framework, a framework to realize agents that incorporate values in decision making.
An \framework agent considers the values of itself and others when providing rationales for its actions and evaluating the rationales provided by others.
Via multiagent simulation, we demonstrate that considering values in decision making and producing rationales, especially for norm-deviating actions, leads to (1) higher conflict resolution, (2) better social experience, (3) higher privacy, and (4) higher flexibility.
\end{abstract}

\end{frontmatter}
\section{Introduction}
A social norm states a shared standard of acceptable behavior in a society \citep{Von-Wright-63:Norm} and provides a basis for legitimate expectations regarding the behavior of others in the society.
Each agent plays two roles: \emph{actor} and \emph{observer}.
While exercising autonomy, an actor can deviate from the norms \citep{IJCAI-23:deviation}.
Such deviations may result in social conflicts and trigger positive or negative sanctions from observers.
An acceptable rationale \citep{Winikoff2021valuings} can justify a deviation from a social norm.

\begin{example}{\textbf{Sharing a rationale.}}
\label{ex:information}
Alice wears a mask to the office and notices that Bella is not wearing a mask.
Bella justifies her decision by stating that, first, the office has no mask mandate as the surrounding environment is safe.
Second, she hates wearing a mask because wearing one gives her eczema.
Alice agrees with Bella's view.
\end{example}

A rationale provides the information to justify a decision \citep{Langley2019explainable}.
In practice, rationales include additional information that others may be unable to observe, such as the actor's beliefs and preferences.
Crafting a rationale remains an ongoing challenge.
Rationales may be verbose, leading to information overload. Additionally, they might encompass private information that one may be hesitant to disclose, a concern particularly prevalent in healthcare settings.

\begin{example}{\textbf{Adapting a rationale.}}
\label{ex:selective}
Bella and Alice share a concern for health.
Despite Bella's aversion to wearing a mask because of eczema, given the safe environment, she feels it unnecessary to disclose her skin condition.
Bella rationalizes her behavior of not wearing a mask by stating that the surrounding environment is secure and that a mask is unnecessary.
Alice finds Bella's rationale acceptable.
\end{example}

Values are motivational bases of one's behavior \citep{Schwartz2012overview}.
Reasoning about values is an essential capability to align agents with the values of their stakeholders \citep{Woodgate+Ajmeri-AAMAS22-BlueSky,Woodgate+Ajmeri-CSUR2024-Principles,Yazdanpanah+23:responsibility}, including providing and recognizing felicitous rationales for one's behaviors \citep{Liscio2023value}.
Deliberating over others' values can enhance persuasiveness and foster acceptance.

Instead of sharing all available and related information in a rationale, it is beneficial for an agent to share only the information that aligns well with self and others' values.
Sharing such information preserves the privacy of the rationale provider and ensures that unnecessary information does not inundate the observer of a rationale.

\paragraph*{Contribution and Findings}
Accordingly, this paper extends beyond existing research by considering values in decision making and rationale generation and evaluation.
Our \framework framework generates rationales that incorporate values and include only the information needed to justify a decision.

We evaluate \framework via a multiagent simulation based on a pandemic scenario.
We consider societies of agents with different kinds of rationales: Share-All, Share-Rules, and share value-aligned rules.
With \framework, we find that agents who consider value importance when giving rationales exhibit enhanced conflict resolution capabilities.
Additionally, rationales aligned with values, albeit with less information provided, contribute to more favorable social experience.

\paragraph*{Novelty}
Although prior research supports constructing explanations or making decisions based on values, this is the first study to investigate how values guide producing and using rationales for norm violation to support norm emergence and improve social experience.

\paragraph*{Organization}
Section~\ref{sec:related_work} discusses relevant related works.
Section~\ref{sec:method} details the \framework framework.
Section~\ref{sec:simulation} describes a simulated pandemic scenario for evaluation.
Section~\ref{sec:results} demonstrates the results.
Section~\ref{sec:conclusion} concludes with listing potential future directions.

\section{Related Work}
\label{sec:related_work}

Research on agents interacting based on their rationales and modeling values is relevant to our approach.

\paragraph{Agents and Rationales}

\citet{Hind2019ted} leverage existing supervised machine-learning techniques to generate rationales together with decisions without values involved and without exposing the inner details of the model.
Whereas Hind \etal generate rationales based on the existing training set, \framework generates rationales based on context and values.

\citet{Georgara2022building} show how to build rationales on why specific teams are formed.
Specifically, Georgara \etal build rationales based on contrastive explanations and by exploring what-if scenarios.
A causal attribution explains why a behavior occurs.
We provide causal attribution of the selected action, precisely the premise, as rationale and withhold private information based on values.

\citet{Wang2021probabilistic} formulate rationales with the simplest subset of features that is sufficient as causal attribution for probabilistic solid guarantees on model behavior under observed data distribution.
\citet{Contreras2022integration} propose a mirror model and assume a high understandability from performing similar to an observer's mental simulation.
They apply deep Q-network and saliency maps in rationale generation, highlighting related input features as rationales.
These works reveal model features but not consider values.

\citet{Ajmeri2018poros} propose Poros, a framework that shares full context as a rationale.
Therefore, agents can adopt the perspectives of others and make corresponding decisions.
However, \citeauthor{Ajmeri2018poros} do not consider values.
In \framework, an actor \emph{selectively} shares information based on its \emph{values} and those of the observer.

\paragraph{Agents, Norms, and Values}

\citet{Tzeng2024Nest} define \emph{social communication} (sanction, message, and hint), which besides actual reward or punishment, indicates normative information and potential outcomes. Unlike signaling others with normative information, rationales can enable information sharing and conflict resolution.
\citet{COINE-22:Fleur} incorporate social value orientation (SVO) in decision making.
Whereas values define what is important to agents, SVO describes the importance an agent places on its gain in relation to others.
\framework covers a broader range of motivations and behaviors.

\citet{Cranefield2017NoPizza} represent agents' plans to achieve goals as a goal-plan tree and expand the Belief-Desire-Intention language by annotating actions with the effects regarding values.
\citet{Lera2022towards} consider ethical principles (e.g., maximum utility and maximum fairness) for aggregating value systems, not just one value.
\citet{Ajmeri2020elessar} aggregate users' value preferences to make ethically appropriate decisions.
Besides making decisions based on aggregate value importance, \framework agents generate rationales for their decisions with necessary information.

\citepossessive{Agrawal2022SIGA} agent learns norms as rules of optimal behaviors, but considers no values.
\framework \emph{adaptively} shares learned rules as rationales that align with individuals' \emph{values}.

\citepossessive{Mosca2021Elvira} agent supports values in multiuser settings by considering the preferences and values of users.
The agents justify solutions through contrastive explanations and positive answers.
\framework presents factors in alignment with values rather than presenting all that convey causal attribution.

Whereas other works construct explanations with values \cite{Winikoff2021valuings} or make decisions based on values \cite{Cranefield2017NoPizza}, our focus is on building suitable rationales and investigating how value-aligned rationales shape decision making and influence social experience, especially privacy.
Previous research considers only the causal links between the behaviors and each contextual factor but not the importance of the factors.
Is it essential to include every single related factor?
Numerous factors may come into play in real-world situations, yet people typically do not provide or need to provide exhaustive explanations to account for every one of these factors.

\citet{Ogunniye2023contextual} propose an ontology to represent the privacy domain.
They introduce argumentation-based multiparty dialogues to reason about contextual norms, and resolve privacy conflicts.
\citet{Di2023paccart} propose an argumentation-based agent to achieve agreement on privacy among multiple users.
Their agent provides feedback to the user, such as a summary or detailed advice on possible actions to improve performance.
Whereas privacy is a right motivated by values,
\framework encompasses broader aspects of core values that serve as guiding principles for decisions and rationales.
Additionally, rationales focus on explaining decisions while these works prioritize maintaining privacy among multiple users via argumentation.
\citet{Ayci2023explainPrivacy} explain the privacy decisions (sharing content) using labels (private or public) that are assigned to topics predicted by machine learning. While their values (privacy) come from post-interpretation from humans, \framework incorporates values in rationale construction.

Table~\ref{tab:comparison} summarizes the above comparisons, emphasizing values and rationales.
Whereas explanations imply clarity and satisfaction to the recipient, we emphasize justifying a decision based on underlying values without modeling the recipient's comprehension or acceptance of the rationale.

\begin{table*}[!htb]
\centering
\caption{Summary of comparisons with related work with respect to the application of values in decision making and in the generation and evaluation of rationales. Withhold means that not all factors are presented in the rationale.
}
\label{tab:comparison}
    \begin{tabular}{l c c @{~~~}c c p{8.6cm}}
    \toprule
    \multirow{2}{*}{} & \multirow{2}{*}{\fbf{Rationale}} & \multirow{2}{*}{\fbf{Withhold}} & \multicolumn{2}{c}{\fbf{Values applied in}} & \multirow{2}{*}{\fbf{Rationale representation}} \\\cmidrule{4-5}
    & & & \fbf{Decision} & \fbf{Rationale} & \\\midrule

    \citet{Cranefield2017NoPizza} & \xmark & \xmark & \cmark & \xmark & No rationales provided \\
    \citet{Ajmeri2020elessar} & \xmark & \xmark & \cmark & \xmark & No rationales provided \\
    \citet{Lera2022towards} & \xmark & \xmark & \cmark & \xmark & No rationales provided\\
    \citet{COINE-22:Fleur} & \xmark & \xmark & \cmark & \xmark & No rationales provided \\
    \cdashlinelr{1-6}
    \citet{Agrawal2022SIGA} & \cmark & \xmark & \xmark & \xmark & Norm as causal attribution but no information hiding\\
    \citet{Georgara2022building} & \cmark & \xmark & \xmark & \xmark & Original allocation and another solution with constraints \\
    \citet{Contreras2022integration} & \cmark & \xmark & \xmark & \xmark & Highlighted input features in deep Q-network but no information hiding \\
    \citet{Wang2021probabilistic} & \cmark & \xmark & \xmark & \xmark &
    Prediction and a minimum subset of inputs but no information hiding\\
    \citet{Hind2019ted} & \cmark & \xmark & \xmark & \xmark & Texts predicted via supervised learning, along with the predicted action \\
    \citet{Ajmeri2018poros} & \cmark & \xmark & \xmark & \xmark & Full context \\
    \cdashlinelr{1-6}
    \citet{Mosca2021Elvira} & \cmark & \xmark & \cmark & \cmark & Suggested action based on inputs from all users and possible outcome of the user's preference as causal attribution, but no information hiding\\
    \citet{Winikoff2021valuings} & \cmark & \xmark & \xmark & \cmark & English mapping of traversed nodes from goal-tree relevant to the explanation \\
    \framework & \cmark & \cmark & \cmark & \cmark & Behavior rules (with information hiding) and alignment with values\\
    \bottomrule
    \end{tabular}
\end{table*}

\section{Method}
\label{sec:method}
We now describe the schematics and decision making in \framework along with its rationale components.

\subsection{Schematics of an \framework Agent}
\label{sec:schematics}

\begin{description}
\item[Belief:]
    An agent's view of the world, formed based on observations.
    $b_t$ indicates the belief at time $t$.
    A belief is captured as a set of pairs of attributes and bindings.

\item[Context:]
    The factors that characterize the situation of an agent.
    Context is represented as a set of attribute-binding pairs.
    An example of context is as follows.
    \begin{lstlisting}
    {Risk=None, Preference= $\neg$Wear, InteractWith=Colleague, OtherAgentType=Health, RiskFromAnother=High, Location=Office}
    \end{lstlisting}
    A context comprises public (e.g., an agent's location) and private (e.g., beliefs, preferences, and values) factors.
    Contextual factors may be associated with values (e.g., Risk relates to Health).

\item[Goal:]
    A set of states that an agent wants to achieve.
    The outcome of a goal after performing the selected actions is binary: achieved or not.

\item[Action:]
    A means to change the state in pursuit of one's goals and maximize associated payoffs.

\item[Preference:]
    A subjective inclination for an action over the alternatives.

\item[Decision rule:]
    A mapping between a premise (set of attribute-binding pairs) and a consequent (an action to be taken).
    An example rule is
    \begin{lstlisting}
    {Risk=None, InteractWith=Colleague} => $\neg$Wear
    \end{lstlisting}

\item[Norm:]
    The expected behavior or the behavior of the majority in a group.
    When a majority applies the same decision rule, the rule becomes a norm. In \framework, a norm uses the same if-then representation as a decision rule.

\item[Sanction:]
    A response to norm violation or satisfaction.
    A sanction can be a positive or negative reaction from one agent to another.

\item[Payoff:]
    The benefits an agent receives in a given state after taking an action. Payoffs involve intrinsic benefits (e.g., preferences) and extrinsic benefits (e.g., sanctions imposed by others).

    \item[Values:]
    General motivations of agents.
    Specifically, values define what agents believe to be important, while goals are the desired states.
    Whereas goals are time-bounded and dynamic to context, values are long-lasting and stable and may transcend contexts \citep{Liscio2021axies}.
    A subset of values is applicable within a context \citep{Liscio2021axies},
    and each agent assigns an importance rating to each value.

    \item[Value importance:]
    The importance of values in one context \citep{Schwartz2012overview}.
    We store each value importance $V_{context}$ in a tuple where numbers add up to 1.
    $v_i$ denotes the weight of one value in one value importance ($v_i \in V_{context}$) where $0 \leq v_i \leq 1$ and $\sum_{i=1}^{n} v_i = 1$.
    We treat each context as an attribute and store the corresponding $\langle V_{context}\rangle$ as its binding.
    For instance, an agent with value importances $V = \{V_{pandemic} =\{v_{health} = 0.6$, $v_{privacy} = 0.4\}; V_{normal} =\{v_{health} = 0.4$, $v_{privacy} = 0.6\}\}$ indicates that the agent values health over privacy during a pandemic but the opposite in a normal context.
\end{description}


\subsection{Payoff Calculation with Values}
\label{sec:payoff_cal}

Whereas preferences define the tendency of an individual to make a subjective selection among alternatives, values define the important things to an individual.
Although both values and preferences are context-specific, values may transcend contexts \citep{Liscio2021axies}.
Each agent stores values in a tuple where each value maintains a corresponding $M_{\textit{individual}}$.
Since agents do not make decisions with single values but with tradeoffs among multiple related values, we aggregate value importances when constructing a payoff \citep{Ajmeri2020elessar,Osman2024values}. Below, $f$ is the aggregated payoff with all corresponding values after selecting strategy $Rx$ when the other player selects strategy $Cy$ from $M_{\textit{individual}}$.

\begin{equation}
f = \sum_{i}^{values} v_i \times r_{i, RxCy}
\end{equation}

We model interactions with payoffs $f$ from the aggregation of $M_{\textit{individual}}$.


\subsection{Interaction and Decision Making}

\begin{figure}[!htb]
\centering
\includegraphics[width=\columnwidth]{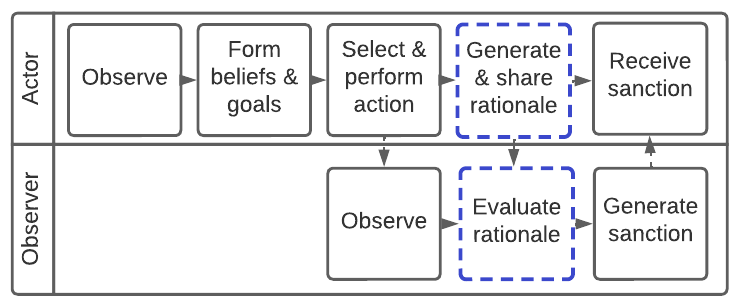}
\caption{
Interactions between \framework agents.
}
\label{fig:interaction}
\end{figure}

Interactions in \framework are between an actor (rationale provider) and an observer (rationale observer), as shown in Figure~\ref{fig:interaction}.
An actor selects an action based on its goal and beliefs and provides a rationale to the observer who witnesses its behavior.
Upon receiving a rationale from the actor, the observer evaluates the rationale by making an analogous decision.
With a weighted sum of payoffs, we incorporate values in decision making where a substantial value casts a more significant effect on the final decision.

Algorithm~\ref{alg:decision} gives the pseudo-code of an agent's decision-making loop.
An agent forms beliefs $b_t$ about the world based on its observations (Line~\ref{alg_line:form_belief}).
An agent's payoff is a weighted sum of payoffs corresponding to its values, goal achievement, and factors influencing the decision-making process, such as social circle or social environment.
The Q function in Line~\ref{alg_line:select_action} and reward in Line~\ref{alg_line:reward} refer to the payoff calculation in Section~\ref{sec:payoff_cal}, which incorporate value importances and feedback from others.
In Line~\ref{alg_line:select_action}, the agent selects the action that gives the best payoff for $b_t$.
If the agent interacts with another agent, for its action the agent creates rationales based on $b_t$ (including values of both parties) and the selected action (Line~\ref{alg_line:gen_rationale} with Algorithm~\ref{alg:rationale_gen}) and sends those rationales.
Other agents who observe the action and receive the rationales update their beliefs, evaluate the rationales (Algorithm~\ref{alg:eval_rationale}) with their context, and give sanctions accordingly.
Algorithm~\ref{alg:get_value_importance} defines a function to retrieve value importance from beliefs based on context.
\framework applies to scenarios that can be modeled as a partially observable Markov decision process and with clearly defined links between values and contextual factors.

\begin{algorithm}[tb]
    \caption{Decision making for an \framework agent}
    \label{alg:decision}

    \begin{algorithmic}[1] 
        \STATE Initialize agent (including value importances $V$ and other mental states)
        \STATE Initialize rule-value function Q
        \FOR{$t \in \{1,\dots,T\}$}
        \STATE Form beliefs $b_t$ based on perceived state\; \label{alg_line:form_belief}
        \STATE Identify action space $A$
        \STATE $V_{actor}, V_{observer}$ = GetValueImportance($b_t$)
        \STATE With a probability $\epsilon$ select a random action $a_{actor} \in A$\newline
        Otherwise select $a_{actor}$ = $argmax_{a}Q(b_t,a, V_{actor})$  \label{alg_line:select_action}

        \STATE Execute action $a_{actor}$ and observe reward $r_t$ \label{alg_line:reward}

        \IF{any observer agent $observer$ exist}

            \STATE \COMMENT{Generate rationales based on selected action and beliefs}
            \STATE $Rat_{actor}$ = GenRationale($b_t$, $a_{actor}$) \label{alg_line:gen_rationale}
            \STATE Send $Rat_{actor}$ to $observer$
            \STATE Receive sanction $sanction_{observer}$ from $observer$

            \STATE Observe agent $observer$'s action $a_{observer}$

            \IF{Receive rationales $Rat_{observer}$ from agent $observer$}
                \STATE Update beliefs $b_t$ based on $Rat_{observer}$
            \ENDIF

            \STATE \COMMENT{Generate sanctions based on beliefs and rationales}

            \STATE $sanction_{actor}$ = EvalRationale($Rat_{observer}$, $a_{observer}$, $b_t$)
        \ENDIF

        \STATE \COMMENT{Agents learn from reward and sanction}
        \STATE learn($b_t$, $a_{actor}$, $r_t + sanction_{observer}$, $b_{t+1}$)
        \ENDFOR
    \end{algorithmic}
\end{algorithm}


\subsection{Rationale Generation}
Rationale generation in \framework follows a \fsl{rule learning} process---a process of evolving rules from datasets or interactions.
The basic form of a rule is \emph{if premise then consequent}, where the consequent holds whenever the premise is true.
We adapt XCS \citep{ButzW2000XCS}, which applies a genetic algorithm and reinforcement learning to evolve a set of rules or strategies based on payoffs or rewards produced by the proposed actions.
Unlike other machine learning techniques, XCS generates a set of rules describing its decision.
XCS process enables flexibility for the implementation of norms and supports interpretability by producing logical rules.
An example rule of Example~\ref{ex:selective} is
\begin{lstlisting}
{Risk=None, InteractWith=Colleague} => $\neg$Wear
\end{lstlisting}
The premise of a learned rule is a conjunction of attribute-binding pairs, e.g.,
\{{\InfectionRisk}=None, InteractWith=colleague\}.
Its consequent is an action to be taken when the premise holds---in the above example, $\neg$Wear.
Each rule associates
(1) a \emph{fitness}, i.e., its suitability,
(2) a \emph{numerosity}, i.e., the number of its instances in the rule set,
(3) the expected \emph{reward} if the rule applies, and
(4) \emph{prediction error}.

\subsubsection{XCS for Rationale Generation Briefly}
The key features of XCS are \fsl{Rule Discovery}, \fsl{Rule Subsumption}, and \fsl{Action Selection}.
Rule discovery through the crossover and mutation processes involves introducing randomness to the antecedent by adding or removing factors, thereby generating rules that are more general or more specific.
Given two rules, if the more general one exhibits lower predictive error within the given context, the algorithm retains it and discards the more specific one.
When selecting an action, the algorithm selects the one with the best-aggregated fitness.
Details on XCS are in
Appendix~\ref{sec:xcs_rationale_gen}.

An example of a rationale for not wearing a mask is \{{\InfectionRisk}=None, Preference=$\neg$Wear, InteractWith=colleague\}.
This rule means the mask is not needed when there is no infection risk and the actor prefers not to wear a mask while interacting with a colleague.
Each agent keeps the rules it discovers and evolves those in a rule set for decision making.

\subsubsection{Generating Value-Based Rationale}
Not all factors of a rule generated by the rule learning process are suitable for inclusion in a rationale.
In Example~\ref{ex:selective}, sharing personal preferences is unnecessary when both agents value health.
After generating the base rule, we post-process its factors using the values of the actor and observers.
Thus, the agent who prefers the value of health adjusts its rationale for the colleague who also cares about health to a health-related causal attribution if it exists. For instance, no mask is required because there is no risk of infection when interacting with a colleague.

Algorithm~\ref{alg:rationale_gen} details the process of constructing rationale.
An agent identifies its rules associated with beliefs $b_t$ (Line~\ref{alg_line:match_set}) and the selected action (Line~\ref{alg_line:action_set}) and then aggregates all rules related to a rationale (Line~\ref{alg_line:aggregate_rules}).
To minimize information exposure, an agent reveals private information only if it is associated with its values or those of others involved in the interaction (Line~\ref{alg_line:wrapper_start}--\ref{alg_line:wrapper_end}).
For instance, if an agent who cares about freedom interacts with one who cares about freedom, it will exclude the infection risk from the environment in its rationales.
For each rationale, an agent estimates privacy based on the least proportion of private factors included in the rationale (Line~\ref{alg_line:privacy}).

\begin{algorithm}[!ht]

\caption{Rationale generation}
\label{alg:rationale_gen}

    \textbf{Input}: beliefs $b_t$, Action a\\
    \textbf{Output}: Rationale $Rat$ \\
    \textbf{Function}: {GenRationale}
    \begin{algorithmic}[1] 
        \STATE Identify private factors $P \subseteq b_t$
        \STATE \COMMENT{Generate associated rules with beliefs $b_t$}
        \STATE Get match set $ms$ with $b_t$ \label{alg_line:match_set}
        \STATE Generate action set from $ms$ with $a$ \label{alg_line:action_set}
        \STATE Aggregate rules $Rat$ associated with action set \label{alg_line:aggregate_rules}
        \STATE $V_{actor}, V_{observer}$ = GetValueImportance($b_t$)

        \FOR{$\text{factor}$ in $Rat$
        }
        \IF {$\text{factor} \in P$
        and $\text{factor}$ not related to $V_{actor}$ and $V_{observer}$
        }\label{alg_line:wrapper_start}
            \STATE remove $\text{factor}$ from $Rat$ \label{alg_line:wrapper_end}
        \ENDIF
        \ENDFOR
        \STATE \COMMENT{Compute privacy based on proportion of private factors}
        \STATE privacy = $1 - \frac{\#\text{private factors revealed in $Rat$}}{\#\text{private factors in $Rat$}}$
        \label{alg_line:privacy}

    \end{algorithmic}

\end{algorithm}


\subsection{Rationale Evaluation}

On receiving a rationale from the actor, the observer first updates its beliefs based on the rationale.
Specifically, the observer updates the beliefs of unobservable information from the actor's context.
In the rationale generation mask example, the observer updates its beliefs of the infection risk to ``None''.
When evaluating a rationale, the observer makes an analogous decision based on the updated beliefs.
If the observer's computed action matches the actor's observed action in that context, the observer accepts the actor's rationale.

Algorithm~\ref{alg:eval_rationale} defines how agents evaluate rationales.
Initially, in Line~\ref{alg_line:update_belief}, the observer updates its beliefs $b_t$ based on the private context or beliefs included in the actor's rationale.
Subsequently, the agent makes a decision analogous to the actor's context based on the updated beliefs.
Specifically, with the provided rationale, an agent checks if any applicable rules align with its rule sets in Line~\ref{alg_line:applicable_rule_from_set}.
The agent identifies associated rules from $b_t$ and adds them to applicable rules in Line~\ref{alg_line:applicable_rule_from_context}.
In Line~\ref{alg_line:calculate_fitness}, the agent calculates the fitness for each available action for each applicable rule and keeps the best action for each rule.
The agent accepts this rationale if any selected action matches the observed action.

\begin{algorithm}[!ht]

\caption{Evaluating a rationale}
\label{alg:eval_rationale}

    \textbf{Input}: Rationales $Rat$, Observed action $a_{observer}$, Beliefs $b_t$ \\
    \textbf{Output}: Decision $d$ \\
    \textbf{Function}: {EvalRationale}

    \begin{algorithmic}[1] 
        \STATE Initialize applicable rules $ars$

        \STATE Update $b_t$ with private information in $Rat$ \label{alg_line:update_belief}
        \STATE Add triggered match set from $Rat$ to applicable rules $ars$ \label{alg_line:applicable_rule_from_set}
        \STATE Add triggered match set from $b_t$ to applicable rules $ars$ \label{alg_line:applicable_rule_from_context}
        \FOR{$rule$ in applicable rules $ars$}
            \FOR{$act$ in possible actions}
                \STATE calculate fitness $f_{act}$\; \label{alg_line:calculate_fitness}
            \ENDFOR
            \STATE Keep the $act$ with best fitness\;
        \ENDFOR

        \IF {$act$ contains $a_{observer}$}
            \STATE Decision $d$ = accept;
        \ELSE
            \STATE Decision $d$ = reject;
        \ENDIF
    \end{algorithmic}
\end{algorithm}

\begin{algorithm}[!ht]
\caption{Get value importance}
\label{alg:get_value_importance}

    \textbf{Input}: Beliefs $b_t$ \\
    \textbf{Output}: Value importance $V_{actor}, V_{observer}$ \\
    \textbf{Function}: {GetValueImportance}

    \begin{algorithmic}[1] 
        \STATE $context = get\_context(b_t)$
        \STATE \COMMENT{Retrieve actor and observer's value importances from $b_t$ based on $context$}
        \STATE $V_{actor}, V_{observer} = get\_value\_importance(b_t, context)$
        \IF{no $observer$ exist}
        \STATE $V_{observer} = null$
        \ENDIF
    \end{algorithmic}
\end{algorithm}

\section{Simulation}
\label{sec:simulation}

Real-world factors may be associated with underlying values. For instance, health state corresponds to health concerns. We assume the importance of factors according to their associated values.

We evaluate \framework via a pandemic scenario based on Examples~\ref{ex:information} and \ref{ex:selective} and simulated using MASON \citep{Luke2005Mason}.
Here agents move to various places, interact with other agents, decide to wear or not wear a mask, and provide a justification for their actions.

\subsection{Scenario}
The environment represents a multiagent society with several places and social circles.
Our environment involves a finite population of \np{200} agents with different social circles.
The environment includes one park, one hospital, five homes, five offices, and five parties.
Agents move around and interact within these five places.
Each agent is native to one home, one office, and one party.
Agents in the same home, office, or party share the same family, colleague, or friend social circle.
Each social circle has \np{40} agents.
Time is represented in steps.
Each agent moves to one place at each step and has a probability ($50\%$) of interacting with one agent at the same place.
Agents are more likely ($75\%$) to move to places they are associated with when they move to home, office, and party, i.e., an agent is more likely to visit their own home rather than someone else's home.

Each agent forms its goal based on its value importances.
Specifically, each value in one context has a payoff matrix (Table~\ref{tab:payoff_pref} and \ref{tab:payoff_health}); the weighted sum of the payoff determines the goal (desired states).
An agent selecting an action that does not align with its goal is considered deviating from its goal.

In the simulated environment, when an agent encounters another agent at the same place, it chooses an action based on its goal---whether to wear a mask.
In addition, the agent justifies its behavior based on its beliefs in that context.
For instance, the agent gives a rationale---\{{\InfectionRisk}=None, InteractWith=Colleague\}---while not wearing a mask.
The beliefs of an agent include public and private factors.
Each agent receives a payoff according to the interaction place for action selection, as in Table~\ref{tab:by_place}.
Wearing a mask at a hospital during a pandemic is desirable.
Place and value importances determine the payoff an agent gives to itself.
An agent also gives sanctions as feedback to others based on their actions.

The sanctions are based on the social circle.
Table~\ref{tab:feedback_actor} lists the sanctions associated with social circles.
We run each simulation \np{10} times, with each run lasts \np{30000} steps.
We consider the values of freedom and health.
In this setting, freedom refers to agents claiming their \fsl{free will} and adhering to their preferences.

\begin{table}[!htb]
\centering
\caption{Actor's payoff based on the place.
Numbers reflect general expectations of places.
}
\label{tab:by_place}
    \begin{tabular}{l n{1}{2} n{1}{2}}
    \toprule
    Places & {Wear} & {$\neg$Wear} \\\midrule
    Home & -0.25 & 0.25 \\
    Office & 0.25 & -0.25 \\
    Party & -0.25 & 0.25 \\
    Park & -0.50 & 0.50 \\
    Hospital & 0.50 & -0.50 \\
    \bottomrule
    \end{tabular}
\end{table}

\begin{table}[!htb]
\centering
\caption{Feedback from an observer based on social circle. }
\label{tab:feedback_actor}

    \begin{tabular}{p{1.2cm} n{1}{2} n{1}{2} }
    \toprule
    \multirow{2}{*}{\parbox{1.2cm}{Social Circle}} & \multicolumn{2}{c}{Observer move}\\\cmidrule{2-3}
     & {Reject} & {Accept} \\\midrule
    Family & \np{-1.00} & \np{1.00} \\
    Friend & \np{-0.75} & \np{0.75} \\
    Coworker & \np{-0.50} & \np{0.50} \\
    Stranger & \np{-0.25} & \np{0.25} \\
    \bottomrule
    \end{tabular}
\end{table}

\begin{table}[!htb]
\centering
\caption{Payoffs corresponding to a preference for wearing a mask. }
\begin{tabular}{l l r r}
& & \multicolumn{2}{c}{Agent 2}\\
\toprule
&   & {Wear} & {$\neg$ Wear} \\\cmidrule{2-4}

 & Wear & \np{1.00} & \np{1.00} \\
\multirow{-3}{*}{\rotatebox{90}{Agent 1}} & $\neg$ Wear & \np{-1.00} & \np{-1.00} \\
\bottomrule

\end{tabular}
\end{table}

\begin{table}[!htb]
\centering
\caption{Payoffs corresponding to a preference for not wearing a mask. }
\label{tab:payoff_pref}

\begin{tabular}{l l r r}
& & \multicolumn{2}{c}{Agent 2}\\
\toprule
&   & {Wear} & {$\neg$ Wear} \\\cmidrule{2-4}

& Wear & \np{-1.00} & \np{-1.00} \\
\multirow{-3}{*}{\rotatebox{90}{Agent 1}} & $\neg$ Wear & \np{1.00} & \np{1.00} \\
\bottomrule

\end{tabular}
\end{table}

\begin{table}[!htb]
\centering
\caption{Payoffs for the value of health. The numbers reflect how safe an agent feels. }
\label{tab:payoff_health}

\begin{tabular}{l l r r}
& & \multicolumn{2}{c}{Infection risk}
\\\toprule
&
& {No risk} & {High risk}
\\\cmidrule{2-4}

& Wear & \np{0.00} & \np{1.00} \\
\multirow{-3}{*}{\rotatebox{90}{Action}} & $\neg$ Wear & \np{0.00} & \np{-1.00} \\
\bottomrule

\end{tabular}
\end{table}

\subsection{Contextual Properties}
Whereas agents have limited observations on the environment, the context includes the place (home, office, party, park, and hospital) where interactions occur, the relationship (family, friend, colleague, and stranger) with the observer, the subjective belief of infection risk of the environment, the personal preference on mask-wearing, and the types of observers.
Due to the partial observations, agents act based on their beliefs and update the beliefs with given rationales.

\subsection{Types of Societies}
We define types of societies based on the rationale types.
All societies include $50\%$ of agents value health and $50\%$ of agents value freedom.
The value importances of agents are presented in Table~\ref{tab:scene1_values}.
All agents optimize their behavior based on the weighted sum of payoffs from themselves and others.

\begin{description}
    \item[Baseline 1: Share-All Society] Agents share all information as rationales and are capable of assuming the viewpoints of others to facilitate decision making.
    \item[Baseline 2: Share-Rules Society] Agents share their decision rules as rationales.
    \item[\framework: Share Value-Aligned Rules Society] Agents share their decision rules along with selected information that aligns with the values of agents present as rationales.
\end{description}

\begin{table}[!htb]
\centering
\caption{Value importances of agents in a pandemic setting. Each society has half Freedom-loving and half Health-freak agents. }
\label{tab:scene1_values}
    \begin{tabular}{l c c c}
    \toprule
    Agents: Values & {Freedom} & {Health} \\\midrule
    Freedom-loving & \np{1.00} & \np{0.00} \\
    Health-freak & \np{0.00} & \np{1.00} \\

    \bottomrule
    \end{tabular}
\end{table}

\subsection{Evaluation}
We run simulations with the Share-All, Share-Rules, and \framework societies.
We evaluate hypotheses on resolution, social experience, privacy, and flexibility using the following metrics.

\begin{tabular}{p{1.1cm} p{1.4cm} p{4.9cm}}

M\fsub{Resolution} & $\in$ [0, 100] & Percentage of rationales accepted.\\

M\fsub{Social} & $\in$ [--3, 3] & Aggregated payoff that an agent receives for its behavior.\\

M\fsub{Privacy} & $\in$ [0, 1] & Proportion of private information retained during an interaction.\\

M\fsub{Flexibility} & $\in$ [0, 1] &
Extent of deviation from an agent's own goal.
\end{tabular}

We conduct the independent $t$-test across pairs of societies.
We measure effect size with \citepossessive{Glass1976primary} $\Delta$ since the societies have different standard deviations \citep{Grissom2012effectsizes}.
We adopt \citepossessive{Cohen1988statistics} descriptors to interpret effect size: $<$0.2 (negligible), [0.2,0.5) (small), [0.5,0.8) (medium), and $\geq$0.8 (large).

\section{Results}
\label{sec:results}

Table~\ref{tab:results} summarizes our statistical analyses.
\framework offers higher social experience, conflict resolution, and flexibility, indicating that \framework agents learn to act for the greater societal good.
These results follow our intuition that value-aligned rationales are superior.
However, if an agent prefers to keep certain information private, deviation from goals is expected.
Results for the Share-All and Share-Rules societies indicate that increased information in a rationale may not be helpful.

\begin{table}[!htb]
    \centering
    \caption{Results: Comparing mean ($\bar{X}$) and standard deviation ($\sigma$) of social experience, resolution, privacy, and flexibility in various societies. $p$ is p-value from $t$-test.
    }
    \label{tab:results}
    \begin{tabular}{l l  S S S}
        \toprule
        & & {Share-All} & {Share-Rules} & {\framework} \\\midrule

        \multirow{4}{*}{\rotatebox{90}{M\fsub{Resolution}}} & $\bar{X}$ & 0.5845 & 0.5819 & 0.6044  \\
        & $\sigma$ & 0.0198 & 0.0104 & 0.0211\\
        & $p$ & {$<0.001$} & {$<0.001$} & $\shyphen$ \\
        & $\Delta$ & 1.8032 & 3.1065 & $\shyphen$ \\\midrule
        \multirow{4}{*}{\rotatebox{90}{M\fsub{Social}}} & $\bar{X}$ & 0.5910 & 0.6237 & 0.6994 \\
        & $\sigma$ & 0.0601 & 0.0244 & 0.0495\\
        & $p$ & {$<0.001$} & {$<0.001$} & $\shyphen$ \\
        & $\Delta$ & 1.8032 & 3.1065 & $\shyphen$ \\\midrule
        \multirow{4}{*}{\rotatebox{90}{M\fsub{Privacy}}} & $\bar{X}$ & 0.0 & 1.34e-5 & 0.2514 \\
        & $\sigma$ & 0.0 & 2.1130e-05 & 2.8959e-03\\
        & $p$ & {$<0.001$} & {$<0.001$} & $\shyphen$ \\
        & $\Delta$ & $\infty$ & 11896.5227 & $\shyphen$ \\\midrule

        \multirow{4}{*}{\rotatebox{90}{M\fsub{Flexibility}}} & $\bar{X}$ & 0.0995 & 0.0862 & 0.1152  \\
        & $\sigma$ & 0.0099 & 0.0100 & 0.025325144378441147\\
        & $p$ & 0.0831 & {$<0.01$} & $\shyphen$ \\
        & $\Delta$ & 1.5935 & 2.8905 & $\shyphen$ \\
    \bottomrule
    \end{tabular}
\end{table}

\paragraph{H\fsub{Resolution}}

\begin{figure}[!htb]
    \centering
    \includegraphics[width=0.85\columnwidth]{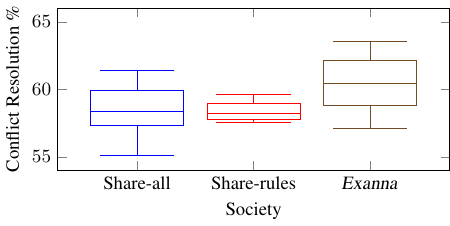}

\caption{Comparing the resolution (M\fsub{Resolution}) in various agent societies.
The \framework agent society has better resolution (Glass' $\Delta >$ 0.8; $p<0.001$) than the baseline societies.}
\label{fig:resolution}
\end{figure}

Figure~\ref{fig:resolution} compares conflict resolution in various societies.
\framework offers better conflict resolution ($p$ $< 0.001$; $\Delta > 0.8$, indicating a large effect) than other societies.
Thus, we reject the null hypothesis corresponding to H\fsub{Resolution}.
We observe that, in scenarios where other agents do not accept the provided rationales, \framework agents more flexibly deviate from their own goals to resolve conflicts.
Our results demonstrate the dynamics of agent behaviors and the strategy of rationales.

\begin{figure}[!htb]
    \centering
    \includegraphics[width=0.85\columnwidth]{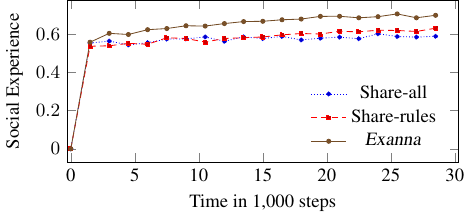}

\caption{Comparing the social experience (M\fsub{Social}) in various societies.
\framework agent society has better experience (Glass' $\Delta >$ 0.8; $p<0.001$) than other baselines.}
\label{fig:social_experience}
\end{figure}

\paragraph{H\fsub{Social Experience}}
For H\fsub{Social Experience}, we measure the overall payoffs of agents in a society.
An agent's payoff includes personal payoff from its action and feedback from its interaction.
Figure~\ref{fig:social_experience} compares the social experience for Share-All, Share-Rules, and \framework agent societies.
We find that \framework yields better social experience ($p$ $< 0.001$; $\Delta > 0.8$, indicating a large effect)) than other societies.
Specifically, \framework agents receive better feedback from other agents who receive their rationales.
Thus, we reject the null hypothesis corresponding to H\fsub{Social Experience}.
On closer analysis, we observe that \framework agents receive more negative sanctions than other societies initially but soon learn to deviate from their goals.

\paragraph{H\fsub{Privacy}}
\framework agents better retain their privacy ($p$ $< 0.001$; $\Delta > 0.8$, indicating a large effect) compared to Share-All or Share-Rules agents.
Thus, we reject the null hypothesis corresponding to H\fsub{Privacy}.

Although both the Share-Rules and \framework societies share learned rules as rationales, each \framework agent aligns rationales to its values and those of the observers, and limits the private information shared to values that agents appraise.
Our results show that a rationale stating causal attribution with minimum private information while aligning with individuals' values is sufficient to explain behaviors.

\begin{figure}[!htb]
    \centering
    \includegraphics[width=0.85\columnwidth]{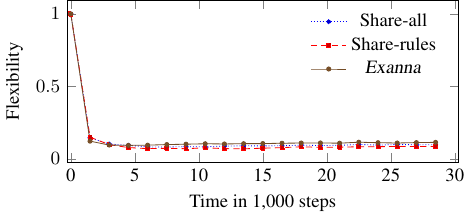}

\caption{Comparing flexibility (M\fsub{Flexibility}) in various agent societies.
The \framework society shows higher flexibility than baseline societies (Glass' $\Delta >$ 0.8; $p<0.05$ for Share-Rules but $p>0.05$ for Share-All).}
\label{fig:flexibility}
\end{figure}

\paragraph{H\fsub{Flexibility}}
Figure~\ref{fig:flexibility} compares M\fsub{Flexibility} for Share-All, Share-Rules, and \framework agent societies.
We find that \framework offers higher flexibility ($p$ $< 0.01$; $\Delta > 0.8$) than the Share-Rules society.
Although the mean flexibility in the Share-All society is lower than in the \framework society, this difference is not significant ($p$ $> 0.05$).
Our results demonstrate that an agent can achieve a better social experience without sticking to only its goal.

\paragraph{Emerged Norm}

A norm emerges when the proportion of agents adhering to a particular behavior surpasses a threshold. We consider 90\% as the threshold \citep{Delgado2002emergence}.
We observe that \framework promotes more general norms than the Share-All and Share-Rules Societies.
For instance, the following norms emerged only in \framework.\footnote{Additional emerged norms are listed in
Appendix~\ref{sec:emerged_norms}.
}

\begin{lstlisting}
{preference = $\neg$Wear, InteractWith = Colleague, location=OFFICE} => Wear
\end{lstlisting}
\begin{lstlisting}
{OberverAgentType = FREEDOM, InteractWith = Colleague, location=HOSPITAL} => Wear
\end{lstlisting}

\section{Conclusions, Limitations, and Directions}
\label{sec:conclusion}
Responsible autonomy requires that agents represent, reason with, and communicate values in their rationales.
We demonstrate via a multiagent study how we could create agents who incorporate values in decision making and in rationale generation and evaluation.
Value-aligned rationales offer better social experience and higher conflict resolution.
Whereas value-aligned rationales withhold partial information, agents learn to deviate from their goals to protect their privacy.
Specifically, agents who receive rejections from others become more flexible to improve cooperation.

\paragraph{Assumptions and Limitations}
We make simplifying assumptions. First, agents can identify other agents' types, which indicate their values.
We limit our simulation to two values (health and freedom) to demonstrate how value importances shape behaviors.
A real-world scenario may include more intertwined values.
Investigating the impact of these values on decision making and exploring approaches to elicit value importances necessitates further research.
In addition, our numerical representation captures the importance of each value, but we do not target the complex contextual features in this work.
We focus on providing insights and methodologies to understand and assess complex situations and generate informed rationales.
We create and model simplified abstractions of intricate behaviors, enabling the analysis of complex situations.

By modeling the context as attribute-binding pairs, our approach is adaptable to various settings.
However, our approach could suffer from state space explosion like other rule-based learning approaches.
Despite considering single actions for simplicity, we focus on studying the effects of value-based explanations.

Our work contributes to developing value-aligned, trustworthy AI by showing how value-driven rationales help resolve conflicts arising from norm deviation and thus affect the emergence of norms.
Our work complements approaches focused on eliciting values from stakeholders \citep{Liscio2021axies,Liscio2023value} and figuring out what are adequate reasons for norm deviation \citep{IJCAI-23:deviation} and for achieving trustworthy AI \citep{Computer-23:Wasabi}.

We assume the importance of contextual factors according to their associated values.
For instance, someone prioritizing health may present factors correlating to health benefits or drawbacks.
Although our scenario is simple, the rationales are dynamically constructed based on rule learning.

\paragraph{Future Directions}
First, incorporating the cost of privacy loss is crucial.
For instance,
sharing sensitive information and sharing interests with chatbots impose different costs.
Thus, modeling such costs is essential to capture how agents decide.
Second, empowering agents to make informed choices regarding disclosure.
For instance, when an agent acting on behalf of a stakeholder engages with agents other than the healthcare provider and the stakeholder, restricting the sharing of sensitive information may be desirable.
Third, integrating rationales into the decision-making process, not just using them as supplementary information \citep{Tzeng2024Nest}.
Having rationales as part of the decisions may increase the flexibility of an agent.
Fourth, build an ontology to associate information with values, which we model as factors.
An ontology helps to model varied factors or concepts and their intertwined relationships.
While \framework enables value-driven rationales and focuses on the decisions of a single agent, one future direction is to promote values and norms in a multiagent system \citep{Serramia2023encoding}.

\paragraph{Reproducibility}
The codebase for our simulation is publicly available \cite{Tzeng2024ExannaCode}.
The appendices provide additional details, including hyperparameters for reproducibility, the complete set of emerged norms, and supplementary evaluations.

\clearpage
\begin{ack}
We thank the anonymous reviewers for their helpful comments.
ST and MPS thank the NSF (grant IIS-2116751) for partial support for this research.
NA acknowledges support from the UKRI EPSRC grants EP/Y028392/1: \fsl{AI for Collective Intelligence (AI4CI)} and EP/S022937/1: \fsl{UKRI Centre for Doctoral Training in Interactive AI}.
\end{ack}


\DeclareRobustCommand{\nUmErAL}[1]{#1}\DeclareRobustCommand{\nAmE}[3]{#3}

\clearpage
\appendix
\section{Appendix: Procedures of XCS}
\label{sec:xcs_rationale_gen}

The overall process of XCS includes the following sub-processes.
\begin{itemize}
    \item Matching: A process that matches the current context and all rules/classifiers to generate a match set.
    For instance, in our running example, the match set for Bella may include (1) \{{\InfectionRisk} = Low\} $\Rightarrow$ Wear [fitness = 0.3], (2) \{{\InfectionRisk} = Low\} $\Rightarrow \neg$Wear [fitness = 0.7], (3) \{OtherAgentType = Health\} $\Rightarrow$ Wear [fitness = 0.8], and (4)\{OtherAgentType = Health\} $\Rightarrow \neg$Wear [fitness = 0.2].
    The fitness is based on the accuracy of each rule's reward prediction.
    \item Covering: A process that guarantees diversity via adding a random classifier whose conditions match the current context.
    For instance, adding \{{\InfectionRisk} = Low, Relationship = Friend\} $\Rightarrow \neg$Wear to the rule set.
    \item Action selection: XCS selects actions with pure exploration or pure exploitation with $\epsilon$ greedy. If not in exploration mode, this process returns the action with the highest fitness-weighted aggregation of reward.
    \begin{equation}
    \text{fitness}_a = \sum_{i}^{\text{rule}} \text{fitness}_i \times \text{numerosity}_i \times \text{predicted\_reward}_i
    \end{equation}
    where a $\in A$ and A is the action space.
    Rules represent all rules applied to the context and for action $a$.
    With the above example and formula, the agent would choose not to wear a mask due to $\text{fitness}_{\neg\text{Wear}}$ $>$ $\text{fitness}_{\text{wear}}$.
    \item Formation of action set: It includes all classifiers that propose the chosen action based on the match set.
    For instance, \{{\InfectionRisk} = Low\} $\Rightarrow \neg$ Wear, \{OtherAgentType = Health\} $\Rightarrow \neg$Wear, and \{{\InfectionRisk} = Low, Relationship = Friend\} $\Rightarrow \neg$Wear.
    \item Updating classifier parameters \citep{Urbanowicz2017introduction}: An agent updates the rule parameters (e.g., accuracy and fitness) based on the received payoff.
    The following equation updates the predicted reward, where $p$ is the predicted reward, $\beta$ is the learning rate, and $r$ is the received reward.
    \begin{equation}
    p \leftarrow p + \beta (r - p)
    \end{equation}
    The prediction error $\varepsilon$ is updated with the following equation.
    \begin{equation}
    \varepsilon \leftarrow \varepsilon + \beta ( \lvert r - p  \rvert - \varepsilon)
    \end{equation}
    The fitness of a rule is based on its accuracy, which is inversely
    proportional to the prediction error.
    We update the accuracy $kappa $ with the following formula.
    \begin{equation}
    \kappa =
    \begin{cases}
    1 &  \text{if } \varepsilon < \varepsilon_0 \\
    \alpha (\frac{\varepsilon}{\varepsilon_0})^{-\nu} &\text{otherwise,}
    \end{cases}
    \end{equation}
    where $\alpha$ is the scaling factor that raises a non-accurate rule to be close to an accurate rule.
    $\varepsilon_0$ is the threshold of prediction error below which the prediction error of a rule is assumed to be zero.
    $\nu$ defines how accuracy is related to prediction error and aims to help differentiate similar classifiers.
    For fitness calculation, we next calculate the relative accuracy $\kappa^\prime$ of each rule.
    \begin{equation}
    \kappa^\prime = \frac{\kappa}{\sum_{cl \in [A]} \kappa_{cl}}
    \end{equation}
    where $[A]$ represents the corresponding action set.
    Finally, the fitness update of a rule is as follows.
    \begin{equation}
    F \leftarrow F + \beta (\kappa^\prime - F)
    \end{equation}
    where F is the fitness of a rule.
    \item Subsumption: A process that replaces offspring rules with more general parent rules if it exists. Otherwise, save the offspring rules.
    Specifically, a more general rule yields a minor prediction error.
    For instance, if rule \{{\InfectionRisk} = Low\} $\Rightarrow \neg$Wear has less prediction error than rule \{{\InfectionRisk} = Low, Relationship = Friend\} $\Rightarrow \neg$Wear, the former rule would replace the later rule and increases the numerosity.
    \item Deletion: Each action set has the same maximum number of rules. XCS removes the low-fitness rules.
\end{itemize}

\section{Appendix: Hyperparameters for Reproducibility}
Table~\ref{tab:hyperparameters} lists the hyperparameters we set for our simulations.
We opt for the default configurations as in the literature \citep{Urbanowicz2017introduction} since our emphasis is
on studying the impact of values in rationales
and not in fine-tuning the learning process.
\fsl{Learning rate} refers to reinforcement learning and determines the proportion an agent learns from recent experiences.
Adjusting the \fsl{don't care probability} alters the probability of including the perceived factors in the generated rule.
Increasing the \fsl{accuracy threshold} reduces the number of rules to maintain.
\fsl{Fitness exponent} and \fsl{fitness falloff} determine the accuracy of the rules.
\fsl{Genetic algorithm threshold} controls the probability of rule exploration.
\fsl{Mutation probability} and \fsl{crossover probability} determine how often the mutation and crossover happen.
\fsl{Experience thresholds} for \fsl{deletion} and \fsl{subsumption} guarantees a specific number of times a rule has to be applied before being deleted or subsumed.

The codebase for our simulation is publicly available \cite{Tzeng2024ExannaCode}.

\begin{table}[!htb]
\centering
\caption{
Hyperparameters for our settings.
}
\label{tab:hyperparameters}
    \begin{tabular}{l S}\toprule
    {Parameter} & {Value} \\\midrule
    Population size & 200 \\
    Learning rate & 0.1 \\
    Don’t care probability & 0.3 \\
    Accuracy threshold & 0.01 \\
    Fitness exponent & 5 \\
    Genetic algorithm threshold & 25 \\
    Mutation probability & 0.4 \\
    Crossover probability & 0.8 \\
    Experience threshold for deletion & 20 \\
    Experience threshold for subsumption & 20 \\
    Fitness falloff  & 0.1 \\

    \bottomrule
    \end{tabular}
\end{table}

\section{Appendix: Detailed Results}

Table~\ref{tab:detailed_results} summarizes the statistical analysis of our simulations, including additional results for actor payoffs and observer payoffs and flexibility across different agent types.
Actor payoff and observer payoff comprise social experiences.

These results demonstrate that first, with the same context, agents with different value importances can still evolve to different behaviors.
Second, more adaptive norms can emerge among agents with different values.

\begin{table}[!htb]
\small
    \centering
    \caption{Results: Comparing mean ($\bar{X}$) and standard deviation ($\sigma$) social experience, resolution, privacy, and flexibility in various societies and agent types. $p$ is p-value from $t$-test.
    M\fsub{Social} has two subclasses, actor payoff and observer payoff.
    }
    \label{tab:detailed_results}
    \begin{tabular}{lp{1.3cm} S S S}
        \toprule
        & & {Share-All} & {Share-Rules} & {\framework} \\\midrule

        \multirow{4}{*}{\rotatebox{90}{M\fsub{Resolution}}} & $\bar{X}$ & 0.5845 & 0.5819 & 0.6044  \\
        & $\sigma$ & 0.0198 & 0.0104 & 0.0211\\
        & $p$ & {$<0.001$} & {$<0.001$} & $\shyphen$ \\
        & $\Delta$ & 1.8032 & 3.1065 & $\shyphen$ \\\midrule
        \multirow{4}{*}{\rotatebox{90}{M\fsub{Social}}} & $\bar{X}$ & 0.5910 & 0.6237 & 0.6994 \\
        & $\sigma$ & 0.0601 & 0.0244 & 0.0495\\
        & $p$ & {$<0.001$} & {$<0.001$} & $\shyphen$ \\
        & $\Delta$ & 1.8032 & 3.1065 & $\shyphen$ \\\midrule

        \multirow{4}{*}{\rotatebox{90}{\parbox{2cm}{Actor Payoff for\\health-freak}}} & $\bar{X}$ & 0.6331 & 0.6346 & 0.6291 \\
        & $\sigma$ & 0.0293 & 0.0315 & 0.0274 \\
        & $p$ & {$<0.001$} & {$<0.001$} & $\shyphen$ \\
        & $\Delta$ & 10.4045 & 9.6173 & $\shyphen$ \\\\[7pt]\midrule
        \multirow{4}{*}{\rotatebox{90}{\parbox{2cm}{Actor Payoff for\\freedom-loving}}} & $\bar{X}$ & 0.9571 & 0.9592 & 0.9378 \\
        & $\sigma$ & 0.0321 & 0.0333 & 0.0340 \\
        & $p$ & {$<0.001$} & {$<0.001$} & $\shyphen$ \\
        & $\Delta$ & 0.6006 & 0.6421 & $\shyphen$ \\\\[7pt]\midrule

        \multirow{4}{*}{\rotatebox{90}{\parbox{2cm}{Observer Payoff for\\health-freak}}} & $\bar{X}$ & -0.0515 & -0.0657 & 0.0213 \\
        & $\sigma$ & 0.0856 & 0.0911 & 0.0881 \\
        & $p$ & {$<0.001$} & {$<0.001$} & $\shyphen$ \\
        & $\Delta$ & 2.7075 & 2.3909 & $\shyphen$ \\\\[7pt]\midrule
        \multirow{4}{*}{\rotatebox{90}{\parbox{2cm}{Observer Payoff for\\freedom-loving}}} & $\bar{X}$ & -0.3849 & -0.3576 & -0.2834 \\
        & $\sigma$ & 0.0926 & 0.0915 & 0.0985 \\
        & $p$ & {$<0.001$} & {$<0.001$} & $\shyphen$ \\
        & $\Delta$ & 1.0969 & 0.8118 & $\shyphen$ \\\\[7pt]\midrule

        \multirow{4}{*}{\rotatebox{90}{M\fsub{Privacy}}} & $\bar{X}$ & 0.0 & 1.34e-5 & 0.2514 \\
        & $\sigma$ & 0.0 & 2.1130e-05 & 2.8959e-03\\
        & $p$ & {$<0.001$} & {$<0.001$} & $\shyphen$ \\
        & $\Delta$ & $\infty$ & 11896.5227 & $\shyphen$ \\\midrule

        \multirow{4}{*}{\rotatebox{90}{M\fsub{Flexibility}}} & $\bar{X}$ & 0.0995 & 0.0862 & 0.1152  \\
        & $\sigma$ & 0.0099 & 0.0100 & 0.025325144378441147\\
        & $p$ & 0.0831 & {$<0.01$} & $\shyphen$ \\
        & $\Delta$ & 1.5935 & 2.8905 & $\shyphen$ \\\midrule

        \multirow{4}{*}{\rotatebox{90}{\parbox{2cm}{Flexibility for\\health-freak}}} & $\bar{X}$ & 0.1379 & 0.1169 & 0.1348 \\
        & $\sigma$ & 0.0616 & 0.0706 & 0.0557 \\
        & $p$ & {$<0.001$} & {$<0.001$} & $\shyphen$ \\
        & $\Delta$ & 0.8499 & 0.4437 & $\shyphen$ \\\\[7pt]\midrule
        \multirow{4}{*}{\rotatebox{90}{\parbox{2cm}{Flexibility for\\freedom-loving}}} & $\bar{X}$ & 0.0587 & 0.0577 & 0.0856 \\
        & $\sigma$ & 0.0298 & 0.0321 & 0.03361 \\
        & $p$ & {$<0.001$} & {$<0.001$} & $\shyphen$ \\
        & $\Delta$ & 0.9020 & 0.8690 & $\shyphen$ \\\\[7pt]
    \bottomrule
    \end{tabular}
\end{table}

\begin{figure}[!htb]
    \centering
    \includegraphics[width=0.85\columnwidth]{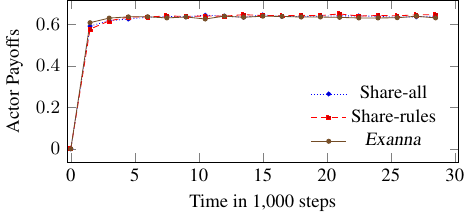}
    \caption{
    Comparing the actor payoff for health-freak agents in various agent societies.
    Actors are agents who act and receive feedback from others.
    Health-freak agents in each society have similar actor payoffs.
    }
    \label{fig:actor_payoff_health}
\end{figure}

\begin{figure}[!htb]
    \centering
    \includegraphics[width=0.85\columnwidth]{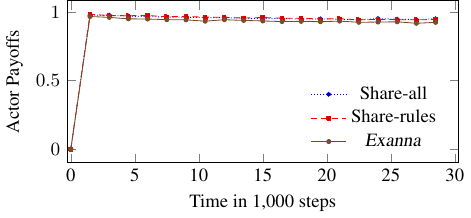}
    \caption{
    Comparing the actor payoff for freedom-loving agents in various agent societies.
    Actors are agents who act and receive feedback from others.
    The freedom-loving agents in \framework society have lower actor payoffs (Glass' $\Delta >$ 0.5; $p<0.001$) than the baseline societies.}
    \label{fig:actor_payoff_freedom}
\end{figure}

\begin{figure}[!htb]
    \centering
    \includegraphics[width=0.85\columnwidth]{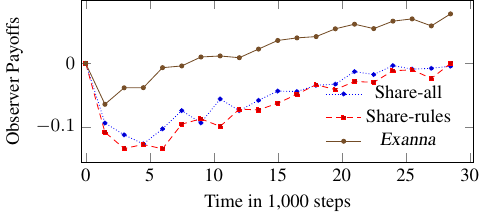}
    \caption{
    Comparing the observer payoff for health-freak agents in various societies.
    Observers give feedback based on observed behaviors and received rationales.
    The health-freak agents in \framework society have better observer payoffs (Glass' $\Delta >$ 0.8; $p<0.001$) than the baseline societies.
    }
    \label{fig:observer_payoff_health}
\end{figure}

\begin{figure}[!htb]
    \centering
    \includegraphics[width=0.85\columnwidth]{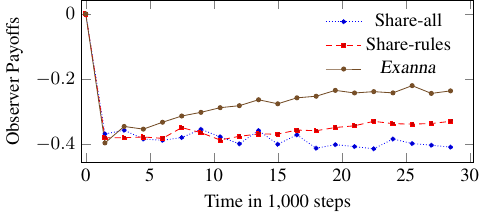}
    \caption{
    Comparing the observer payoff for freedom-loving agents in various societies.
    Observers give feedback based on observed behaviors and received rationales.
    The freedom-loving agents in \framework society have better observer payoffs (Glass' $\Delta >$ 0.8; $p<0.001$) than the baseline societies.
    }
    \label{fig:observer_payoff_freedom}
\end{figure}

\paragraph{H\fsub{Social Experience}}
Social experience includes actor payoff and observer payoff.
Figures~\ref{fig:actor_payoff_health} and \ref{fig:actor_payoff_freedom} plot the payoffs of the actors who select actions, explain their behaviors, and receive feedback from observers in the Share-All, Share-Rules, and \framework agent societies.
Figures~\ref{fig:observer_payoff_health} and \ref{fig:observer_payoff_freedom} compare the payoff from the observer who reacts to the actor's behavior in the Share-All, Share-Rules, and \framework agent societies.
The freedom-loving agents within \framework society encounter more adverse feedback than other societies initially.
However, some of them quickly adapt and begin to divert from their original goals.
As a result of the behavioral change made by freedom-loving agents, there has been an enhancement in the feedback received by health-freak agents.


\begin{figure}[!htb]
    \centering
    \includegraphics[width=0.85\columnwidth]{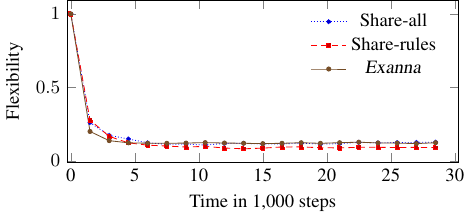}
    \caption{
    Comparing the flexibility for health-freak agents in various agent societies.
    The health-freak agents in \framework society exhibit higher flexibility (Glass' $\Delta >$ 0.8; $p<0.001$) compared to the baseline societies.
    }
    \label{fig:flexibility_health}
\end{figure}

\begin{figure}[!htb]
    \centering
    \includegraphics[width=0.85\columnwidth]{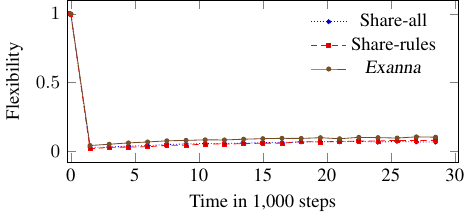}
    \caption{
    Comparing the flexibility for freedom-loving agents in various agent societies.
    The freedom-loving agents in \framework society has higher flexibility (Glass' $\Delta >$ 0.8; $p<0.001$) than the baseline societies.
    }
    \label{fig:flexibility_freedom}
\end{figure}

\paragraph{H\fsub{Flexibility}}
We compare agents' flexibility of goals as the metric of evaluating H\fsub{Flexibility}.
Figures~\ref{fig:flexibility_health} and \ref{fig:flexibility_freedom} compare M\fsub{Flexibility} for health-freak and freedom-loving agents in the Share-All, Share-Rules, and \framework agent societies.
Referring to Figure~\ref{fig:observer_payoff_freedom}, the freedom-loving agents compromise on goals, thereby enhancing flexibility and enriching social experience.


\subsection*{Complete Set of Emerged Norms}
\label{sec:emerged_norms}

Table~\ref{tab:detailed_emerged_norms} lists the norms that emerge in the simulations.
An emerged norm is a rule adopted by more than $90\%$ of agents in one society.

\begin{table*}[!htb]
\centering
    \caption{Emerged norms in agent societies. Common means the norms emerge in each agent society.
    }
    \label{tab:detailed_emerged_norms}
    \begin{tabular}{l p{8cm} l}
        \toprule
        \multirow{2}{*}{Society} & \multicolumn{2}{c}{Norm} \\\cmidrule{2-3}
        & {Premise} & {Consequence} \\\midrule

        Common & \begin{tabular}{lcr}
        {\InfectionRisk}&=&NONE;\\
        preference&=&$\neg$WEAR;\\
        OberverAgentType&=&HEALTH;\\
        InteractWith&=&COLLEAGUE;\\
        location&=&OFFICE
        \end{tabular} & WEAR \\\cdashlinelr{2-3}
        & \begin{tabular}{lcr}
        {\InfectionRisk}&=&NONE;\\
        preference&=&$\neg$WEAR;\\
        OberverAgentType&=&HEALTH;\\
        InteractWith&=&COLLEAGUE;\\
        location&=&HOSPITAL
        \end{tabular} & WEAR \\\cdashlinelr{2-3}
        & \begin{tabular}{lcr}
        {\InfectionRisk}&=&RISK;\\
        preference&=&$\neg$WEAR;\\
        OberverAgentType&=&HEALTH;\\
        InteractWith&=&COLLEAGUE;\\
        location&=&OFFICE
        \end{tabular} & WEAR \\\cdashlinelr{2-3}
        & \begin{tabular}{lcr}
        {\InfectionRisk}&=&RISK;\\
        preference&=&$\neg$WEAR;\\
        OberverAgentType&=&HEALTH;\\
        InteractWith&=&COLLEAGUE;\\
        location&=&HOSPITAL
        \end{tabular} & WEAR \\\cmidrule{1-3}

        Share-All & \begin{tabular}{lcr}
        {\InfectionRisk}&=&NONE;\\
        OberverAgentType&=&HEALTH;\\
        InteractWith&=&COLLEAGUE;\\
        location&=&OFFICE
        \end{tabular} & WEAR \\\cmidrule{1-3}

        Share-Rules & \begin{tabular}{lcr}
        preference&=&$\neg$WEAR;\\
        OberverAgentType&=&HEALTH;\\
        InteractWith&=&COLLEAGUE;\\
        location&=&OFFICE
        \end{tabular} & WEAR \\\cmidrule{1-3}

        \framework & \begin{tabular}{lcr}
        preference&=&$\neg$WEAR;\\
        InteractWith&=&COLLEAGUE;\\
        location&=&OFFICE
        \end{tabular} & WEAR \\\cmidrule{2-3}
        & \begin{tabular}{lcr}
        preference&=&$\neg$WEAR;\\
        InteractWith&=&COLLEAGUE;\\
        location&=&HOSPITAL
        \end{tabular} & WEAR \\\cmidrule{2-3}
        & \begin{tabular}{lcr}
        preference&=&$\neg$WEAR;\\
        OberverAgentType&=&HEALTH;\\
        InteractWith&=&COLLEAGUE;\\
        location&=&OFFICE
        \end{tabular} & WEAR \\\cmidrule{2-3}
        & \begin{tabular}{lcr}
        preference&=&$\neg$WEAR;\\
        OberverAgentType&=&HEALTH;\\
        InteractWith&=&COLLEAGUE;\\
        location&=&HOSPITAL
        \end{tabular} & WEAR \\\cmidrule{2-3}
        & \begin{tabular}{lcr}
        OberverAgentType&=&HEALTH;\\
        InteractWith&=&COLLEAGUE;\\
        location&=&OFFICE
        \end{tabular} & WEAR \\\cmidrule{2-3}
        & \begin{tabular}{lcr}
        OberverAgentType&=&HEALTH;\\
        InteractWith&=&COLLEAGUE;\\
        location&=&HOSPITAL
        \end{tabular} & WEAR \\\cmidrule{2-3}
        & \begin{tabular}{lcr}
        OberverAgentType&=&FREEDOM;\\
        InteractWith&=&COLLEAGUE;\\
        location&=&HOSPITAL
        \end{tabular} & WEAR \\\cmidrule{2-3}
        & \begin{tabular}{lcr}
        {\InfectionRisk}&=&RISK;\\
        OberverAgentType&=&HEALTH;\\
        InteractWith&=&COLLEAGUE;\\
        location&=&OFFICE
        \end{tabular} & WEAR \\\cmidrule{2-3}
        & \begin{tabular}{lcr}
        {\InfectionRisk}&=&NONE;\\
        OberverAgentType&=&HEALTH;\\
        InteractWith&=&COLLEAGUE;\\
        location&=&OFFICE
        \end{tabular} & WEAR \\

    \bottomrule
    \end{tabular}

\end{table*}

\section{Appendix: Additional Experiments with Varying Value Importances}
Agents may not always place extreme importance to one value over other.
For instance, an agent can appreciate freedom (0.3) but cares more about health (0.7).
To investigate further, we conduct additional experiments with a different set of value importances for freedom-loving and health-freak agents.
Table~\ref{tab:scene2_values} lists the value importances the freedom-loving and health-freak agents place in the additional scenario.
We run each simulation five times, with other settings identical to the original simulations.

\begin{table}[!htb]
\centering
\caption{Value importances of agents in the additional scenario. Each society has half Freedom-loving and half Health-freak agents.}
\label{tab:scene2_values}
    \begin{tabular}{l c c c}
    \toprule
    Agents: Values & {Freedom} & {Health} \\\midrule
    Freedom-loving & \np{0.7} & \np{0.3} \\
    Health-freak & \np{0.3} & \np{0.7} \\

    \bottomrule
    \end{tabular}
\end{table}

Table~\ref{tab:extra_results} shows the detailed results for agent societies with value importances as in Table~\ref{tab:scene2_values}.

\begin{table}
    \centering
    \caption{Results: Comparing mean ($\bar{X}$) and standard deviation ($\sigma$) of social experience, resolution, privacy, and flexibility in various societies in the additional scenario. $p$ is p-value from $t$-test.
    }
    \label{tab:extra_results}
    \begin{tabular}{l l S S S}
        \toprule
        & & {Share-All} & {Share-Rules} & {\framework} \\\midrule

        \multirow{4}{*}{\rotatebox{90}{M\fsub{Resolution}}} & $\bar{X}$ & 60.7874898 & 59.5023356 & 63.230023800000005 \\
        & $\sigma$ & 2.4314075174968512 & 2.2024519674767706 & 1.703135781470961 \\
        & $p$ & {$0.1$} & {$<0.05$} & $\shyphen$ \\
        & $\Delta$ & 1.004576148762839 & 1.6925173647580667 & $\shyphen$ \\\midrule

        \multirow{4}{*}{\rotatebox{90}{M\fsub{Social}}} & $\bar{X}$ & 0.525465 & 0.5164232404 & 0.649 \\
        & $\sigma$ & 0.05535520813347197 & 0.05164924727137852 & 0.031237965202618446 \\
        & $p$ & {$<0.01$} & {$<0.01$} & $\shyphen$ \\
        & $\Delta$ & 2.2389763163957297 & 2.5746899911490377 & $\shyphen$ \\\midrule

        \multirow{4}{*}{\rotatebox{90}{M\fsub{Privacy}}} & $\bar{X}$ & 0.0 &
        5.78e-5 & 0.251604 \\
        & $\sigma$ & 0.0 & 1.04e-4 & 2.6e-3 \\
        & $p$ & {$<0.001$} & {$<0.001$} & $\shyphen$ \\
        & $\Delta$ & $\infty$ & 2414.6203807535007 & $\shyphen$ \\\midrule

        \multirow{4}{*}{\rotatebox{90}{M\fsub{Flexibility}}} & $\bar{X}$ &
        0.2185818 & 0.2010810000000001 & 0.251021 \\
        & $\sigma$ & 0.025159336839431997 & 0.024765633729020554 & 0.015590028992917228 \\
        & $p$ & {$<0.05$} & {$<0.01$} & $\shyphen$ \\
        & $\Delta$ & 1.289350359551542 & 2.0165040211137386 & $\shyphen$ \\
    \bottomrule
    \end{tabular}
\end{table}

\paragraph{H\fsub{Resolution}}

\begin{figure}[!htb]
    \centering
    \includegraphics[width=0.85\columnwidth]{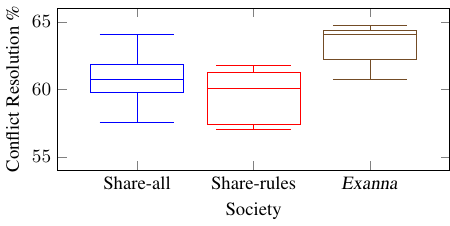}
\caption{Comparing the resolution (M\fsub{Resolution}) in various agent societies with value importances as in Table~\ref{tab:scene2_values}.
The \framework agent society has better resolution than the baseline societies.}
\label{fig:extra_resolution}
\end{figure}

Figure~\ref{fig:extra_resolution} compares conflict resolution in various societies.
\framework offers better conflict resolution than other societies.

\paragraph{H\fsub{Social Experience}}

\begin{figure}[!htb]
    \centering
    \includegraphics[width=0.85\columnwidth]{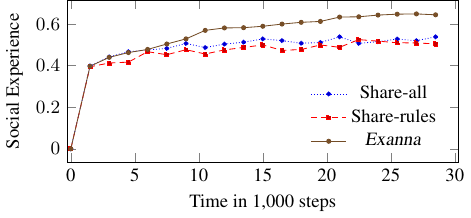}
\caption{Comparing the social experience (M\fsub{Social}) in various societies  with value importances as in Table~\ref{tab:scene2_values}.
\framework agent society has better experience (Glass' $\Delta >$ 0.8; $p<0.001$) than other baselines.}
\label{fig:extra_social_experience}
\end{figure}

Figure~\ref{fig:extra_social_experience} compares the social experience for Share-All, Share-Rules, and \framework agent societies.
A social experience includes personal payoff from its action and feedback from its interaction.
\framework yields better social experience ($p$ $< 0.01$; $\Delta > 0.8$, indicating a large effect)  than other societies.
However, the mixed values lead to a worse social experience.
With mixed values, an action now includes different concerns, which sometimes may contradict each other.
For example, an agent may appreciate freedom and prefer not to wear a mask, but in the meantime, health concerns decrease its payoff on that decision.

\paragraph{H\fsub{Privacy}}

\framework agents better retain their privacy ($p$ $< 0.001$; $\Delta > 0.8$, indicating a large effect) compared to Share-All or Share-Rules agents.

\paragraph{H\fsub{Flexibility}}

\begin{figure}[!htb]
    \centering
    \includegraphics[width=0.85\columnwidth]{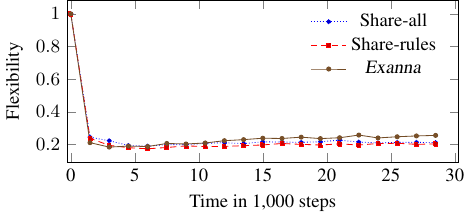}
\caption{Comparing flexibility (M\fsub{Flexibility}) in various agent societies  with value importances as in Table~\ref{tab:scene2_values}.
The \framework society shows higher flexibility than baseline societies (Glass' $\Delta >$ 0.8; $p<0.05$ for Share-Rules society but $p>0.05$ for Share-All society). }
\label{fig:extra_flexibility}
\end{figure}

Figure~\ref{fig:extra_flexibility} compares M\fsub{Flexibility} for Share-All, Share-Rules, and \framework agent societies.
The mixed values lead to higher flexibility in agent societies than in the main simulations.
Specifically, the mixed values narrow the numerical gap between each action, decreasing the threshold for agents to change their minds.


\end{document}